\documentstyle[prl,eqsecnum,aps,epsf]{revtex}
\begin{document}
\draft

\title{Polarization transfer in $^4$He$(\vec{e},e^\prime\vec{p}\,)$$^3$H: is the 
ratio $G_{Ep}/G_{Mp}$ modified in medium ?}
\author{R.\ Schiavilla}
\address{Jefferson Lab, Newport News, Virginia 23606 \\
         and \\
         Department of Physics, Old Dominion University, Norfolk, Virginia 23529, USA}
\author{O.\ Benhar}
\address{Istituto Nazionale di Fisica Nucleare and Dipartimento di Fisica,
Universit\`a \lq\lq La Sapienza\rq\rq, I-00185 Roma, Italy}
\author{A.\ Kievsky, L.E.\ Marcucci, and M.\ Viviani}
\address{Istituto Nazionale di Fisica Nucleare and Dipartimento di Fisica, 
Universit\`a di Pisa, I-56100 Pisa, Italy}
\date{\today}
\maketitle

\begin{abstract}
Polarization observables in the $^4$He$(\vec{e},e^\prime\vec{p}\,)$$^3$H
reaction are calculated using accurate three- and four-nucleon bound-state
wave functions, a realistic model for the nuclear electromagnetic current
operator, and a treatment of final-state-interactions with an optical
potential.  In contrast to earlier studies, no significant discrepancies
are found between theory and experiment both for the ratio of transverse to
longitudinal polarization transfers and for the induced polarization, when
free-nucleon electromagnetic form factors are used in the current operator.
The present results challenge the current interpretation of the experimental
data in terms of medium-modified form factors.
\end{abstract}

The recent measurement, carried out at Jefferson Lab (JLab)~\cite{Strauch03}
(E93-049), of the ratio of transverse, $P_x^\prime$, to longitudinal, $P_z^\prime$,
polarization transfer parameters in the $^4$He$(\vec{e},e^\prime\vec{p}\,)$$^3$H
reaction has generated considerable interest.  In the elastic process
$\vec{e}p\rightarrow e\vec{p}$, the $P_x^\prime/P_z^\prime$ ratio is
proportional to that of electric to magnetic form factors of the proton~\cite{Jones00},
and therefore its measurement in a nucleus by quasi-elastic proton knockout
can shed light, in principle, on the question of whether these form factors
are modified in medium.  However, the answer to this question is obviously
model dependent, since the modification is inferred from a comparison of the
experimental data with theoretical predictions for $(\vec{e},e^\prime \vec{p}\,)$
cross sections {\it in nuclei}.  Therefore, it is crucial for
a proper interpretation of the experimental
data that the theoretical calculations include contributions from final
state interactions (FSI) between the knocked out proton and residual system,
as well as from many-body terms in the nuclear electromagnetic current and
from correlation effects in both the initial and final bound nuclear clusters. 
At issue then is whether these contributions have so far been accounted for and
reliably estimated.  For example, the studies of Refs.~\cite{Udias99,Lava04}, based
on relativistic mean field theory, ignore correlation effects in the bound
state wave functions and many-body terms in the electromagnetic operator.
Furthermore, FSI are treated with a relativistic optical potential
in the work of Udias and collaborators~\cite{Udias99}, in which the
contributions associated with charge exchange processes are neglected---they will
turn out to play an important role in the reaction under consideration, see below.
In the work of the Ghent group~\cite{Lava04}, FSI are described in a Glauber
framework, which may not be reliable at the low end of the $Q^2$-range covered
by E93-049, since the ejected proton energies are too low.  In  addition and
more importantly, the charge-exchange mechanism referred to earlier is also not
included in this study---indeed, it is not obvious how to incorporate it
within the context of a Glauber approach.  Lastly, Laget's calculations~\cite{Laget04},
a full account of which is yet to be published, treat FSI by retaining
S, P, and D waves in the nucleon-nucleon ($N$$N$) scattering amplitude
at low energy, and by using a standard parameterization of the latter
in terms of a central term at higher energies.  Charge-exchange as well
as spin-dependent effects beyond those implicit in the use of the low-energy
$N$$N$ amplitudes, are neglected.  Two-body terms in the current operator
are shown to lead to a quenching of $\simeq$ 2--2.5 \% in the ratio
$P_x^\prime/P_z^\prime$ relative to that in Plane-Wave-Impulse-Approximation
(PWIA)---note, however, that in Ref.~\cite{Strauch03} no quenching is reported 
in the result at $Q^2$=0.5 (GeV/c)$^2$ (the only one shown) for the calculation
from the same author.  Approximations are made in the numerical evaluations of
the loop integrals occurring in Laget's diagrammatic approach.

The present study is based on variational wave functions for the bound
three- and four-nucleon systems, derived from a realistic Hamiltonian
consisting of the Argonne $v_{18}$ two-nucleon~\cite{Wiringa95} and
Urbana-IX three-nucleon~\cite{Pudliner95} interactions (AV18/UIX) with the
hyperspherical-harmonics (HH) technique, with ($^3$He/$^3$H)~\cite{Kievsky93}
and without ($^4$He)~\cite{Viviani04} the inclusion of pair correlations.
The high accuracy of the HH wave functions is well documented~\cite{Nogga03},
as is the quality of the AV18/UIX Hamiltonian in successfully and quantitatively
accounting for a wide variety of three- and four-nucleon bound-state
properties and reactions, ranging from binding energies,
charge radii, and elastic form factors~\cite{Nogga03,Marcucci98,Carlson98}
to low-energy radiative and weak capture cross sections and
polarization observables~\cite{Marcucci04}, to the quasi-elastic response in
inclusive $(e,e^\prime)$ scattering at intermediate energies~\cite{Carlson00}. 

The polarization transfer measurement in the JLab experiment
E93-049 was performed in a quasi-elastic regime: the momentum
of the recoiling $^3$H nucleus was kept close to zero.  The proton lab kinetic
energies were (0.29,0.55,0.88,1.42) GeV  for the $Q^2$ values
(0.5,1.0,1.6,2.6) (GeV/c)$^2$, respectively.  These energies
are obviously beyond the range of applicability of $N$$N$
interaction models, such as the AV18, which are constrained to reproduce $N$$N$
elastic scattering data up to the pion production threshold.  At higher energies,
$N$$N$ scattering becomes strongly absorptive with the opening of particle production
channels.  Indeed, the $p$$p$ inelastic cross section at 0.5 GeV increases abruptly
from about 2 mb to 30 mb, and remains essentially constant for energies up to several
hundred GeV~\cite{Lechanoine93}.

In view of these considerations, FSI in the $p\, ^3$H scattering state are
described in the present work via an optical potential~\cite{vanOers82,Schiavilla90}.
Of course, this approximation has limitations as to the energy range where it is expected
to be valid, see discussion below.  The $p\, ^3$H wave function is then written as

\[
\psi^{(-)}_{{\bf k}\sigma;\sigma_3}(p+^3\!{\rm H})=
\frac{1}{\sqrt{4}} \sum_{\rm perm} (-)^{\rm perm}
\Big[ \eta_{{\bf k}\sigma}^{(-)}(i;p) \phi_{\sigma_3}(jkl;^3\!{\rm H}) 
      +\eta_{{\bf k}\sigma}^{(-)}(i;n) \phi_{\sigma_3}(jkl;^3\!{\rm He}) \Big] \ ,
\]
where $\sigma$ and $\sigma_3$ are the spectator nucleon and bound cluster
spin projections, ${\bf k}$ is their relative momentum, and the sum
over permutations ensures the antisymmetry of the wave function
$\psi^{(-)}$.  The spectator wave functions $\eta(i;p/n)$
are given by the linear combinations $[\eta(i;T=1)$$+/-$$\eta(i;T=0) ]/2$,
where $T$=0,1 denotes the total isospin of the 1+3 clusters.  The
latter are taken to be the scattering solutions of a Schr\"odinger
equation containing a complex, energy-dependent optical potential
of the form

\[
v^{\rm opt}_T(T_{\rm rel})= [ v^c(r;T_{\rm rel})+(4T-3)v^{c\tau}(r;T_{\rm rel})] 
                           +[ v^b(r;T_{\rm rel})+(4T-3)v^{b\tau}(r;T_{\rm rel})]\,
{\bf l}\cdot {\bf s} \ ,
\]
where $T_{\rm rel}$ is the relative energy between clusters $i$ and $j$$k$$l$,
and ${\bf l}$ and ${\bf s}$ are the orbital and spin angular momenta of nucleon $i$,
respectively.  The imaginary part of $v^{\rm opt}_T$ accounts for the loss of
flux in the $p\, ^3$H and $n\, ^3$He states due to their coupling
to the $dd$, three- and four-body breakup channels of $^4$He.
Note that the $n$+$^3$He component in the scattering wave function
$\psi^{(-)}(p+^3\!{\rm H})$ vanishes unless the isospin-dependent
(charge-exchange) terms in $v^{\rm opt}$ are included.  In the results
presented below, all partial waves are retained in the expansion of $\eta(i;T)$,
with full account of interaction effects in those with relative orbital angular
momentum $l \leq 17$.  It has been explicitly verified that the numerical importance
of FSI in higher partial waves is negligible.

The central $v^c$ and $v^{c\tau}$, and spin-orbit $v^b$ and $v^{b\tau}$
terms have standard Woods-Saxon and Thomas functional forms.  The parameters
of $v^c$, $v^{c\tau}$, and $v^b$ were determined by fitting $p+^3{\rm H}$
elastic cross section data in the lab energy range
$T_{\rm lab}$=(160--600) MeV, and $p+^3{\rm H} \rightarrow n+^3{\rm He}$
charge-exchange cross section data at $T_{\rm lab}$=57 MeV and 156 MeV
(see Refs.~\cite{vanOers82,Schiavilla90} for a listing of their values).
The charge-exchange spin-orbit term is taken to be purely real, with
a depth parameter depending logarithmically on $T_{\rm lab}$,
$15.0-1.5\, {\rm log}[T_{\rm lab}({\rm MeV})]$ in MeV, and with radius and
diffuseness having the values 1.2 fm and 0.15 fm, respectively.
The isospin-independent and isospin-dependent spin-orbit terms of
$v_T^{\rm opt}$ are not well constrained by the data, since
these consist exclusively of differential cross
sections~\cite{vanOers82,Langevin70,Darves72}.  However, they
significantly affect the induced polarization $P_y$ measured in the
$^4$He$(\vec{e},e^\prime \vec{p}\,)$$^3$H, as will be shown below.

The nuclear electromagnetic current includes one- and two-body terms.
The one-body current and charge operators have the form recently
derived by Jeschonnek and Donnelly~\cite{Jeschonnek98} (specifically,
Eqs. (23) and (25) of Ref.~\cite{Jeschonnek98}) from an expansion
of the covariant single-nucleon current, in which only quadratic and
higher order terms are neglected in its dependence on the initial
nucleon momentum.  This form of the one-body currents is well
suited for dealing with processes in which the energy transfer
may be large (i.e., the ratio of four- to three-momentum transfer
$(Q/q)^2$ is not close to one) and the initial momentum of the
struck nucleon is small.  Thus, its use is certainly justified in the
quasi-elastic kinematics of the E93-049 experiment under consideration.
In the limit $(Q/q)^2 \simeq 1$ one recovers the standard expressions
for the impulse-approximation currents (including the spin-orbit
correction to the charge operator).

The two-body charge and current operators consist of a
\lq\lq model-independent\rq\rq part, that is constructed
from the $N$$N$ interaction (the AV18 in the present case),
and a \lq\lq model-dependent\rq\rq one, associated with the
excitation of intermediate $\Delta$ resonances and the $\rho\pi\gamma$
and $\omega\pi\gamma$ transition mechanisms (for a review, see
Ref.~\cite{Carlson98} and references therein).

Finally, the H\"ohler parameterization~\cite{Hohler76} is used for the
electromagnetic form factors of the nucleon, except at the highest $Q^2$ values
of 1.6 (GeV/c)$^2$ and 2.6 (GeV/c)$^2$, for which the proton electric and
magnetic form factors are taken from the parameterization obtained in
Ref.~\cite{Brash02} by fitting $G_{Mp}$ data and the ratio $G_{Ep}/G_{Mp}$
recently measured at JLab~\cite{Jones00}.  Incidentally, E93-049
also reported measurements of the polarization transfer ratio
on hydrogen in the same kinematics as for helium~\cite{Strauch03}.
The results for $(P_x^\prime/P_z^\prime)_{^1{\rm H}}$ are in agreement
with the H\"ohler-Brash parameterization, except at $Q^2$=1.6 (GeV/c)$^2$:
Strauch {\it et al.} measure $-0.395 \pm 0.013$ while the Brash fit, based
on Ref.~\cite{Jones00}, gives $-0.415$.  However, this difference is
due to finite detector acceptances~\cite{Strauch04}.

\begin{figure}[bth]
\let\picnaturalsize=N
\def\picsize{3in}
\def\picfilenamea{ratio.eps}
\epsfbox{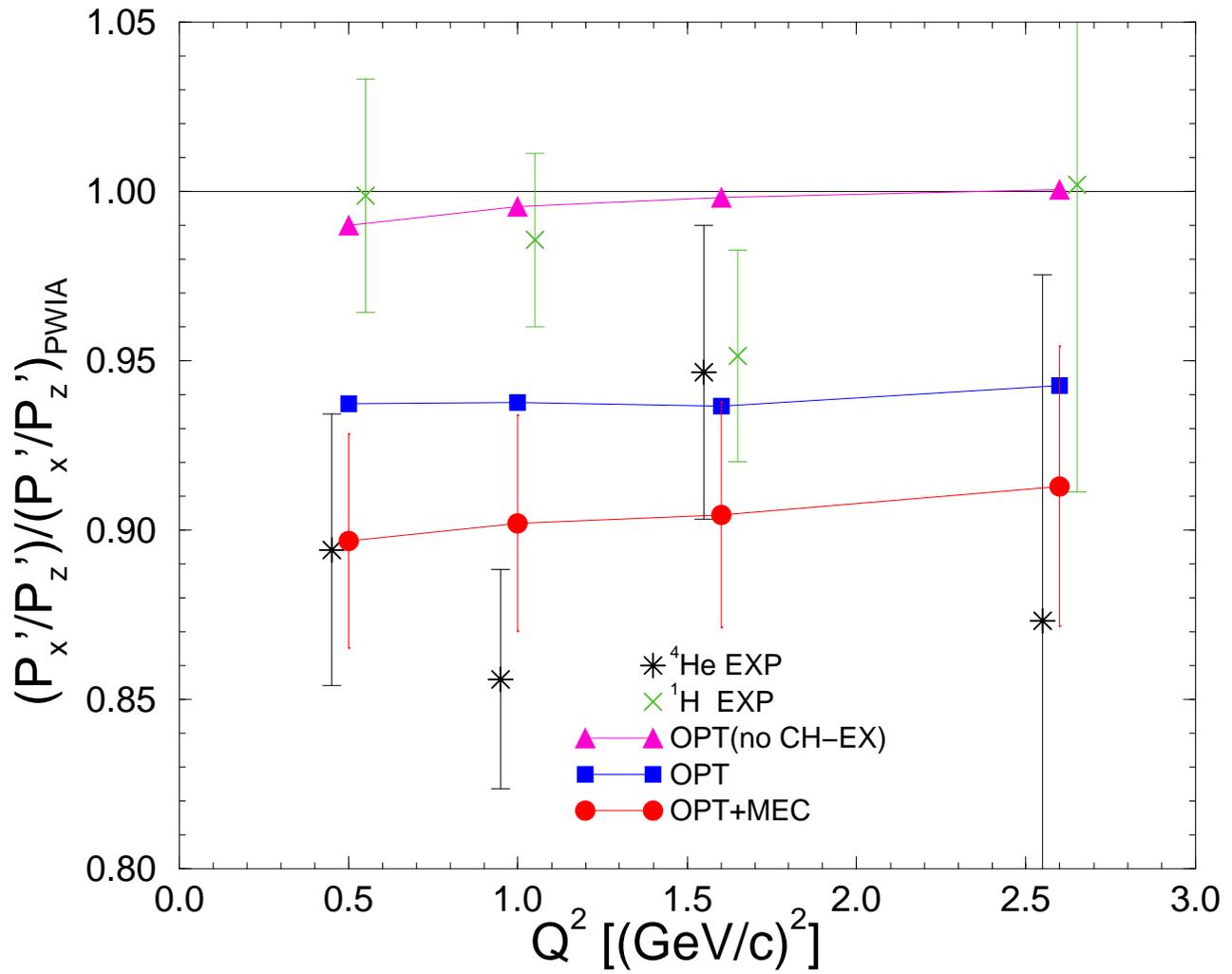}
\caption{The super-ratios measured for $^4$He are compared to theoretical predictions,
obtained in various approximation schemes (see text for an explanation of the notation).
The solid lines are to guide the eye only.  Also shown are the super-ratios measured for $^1$H.
Note that the $^4$He ($^1$H) data have been shifted to the left (right) by 0.05 (GeV/c)$^2$
in order to reduce clutter.}
\label{fig:ratio}
\end{figure}

\begin{figure}[bth]
\let\picnaturalsize=N
\def\picsize{3in}
\def\picfilenamea{py.eps}
\epsfbox{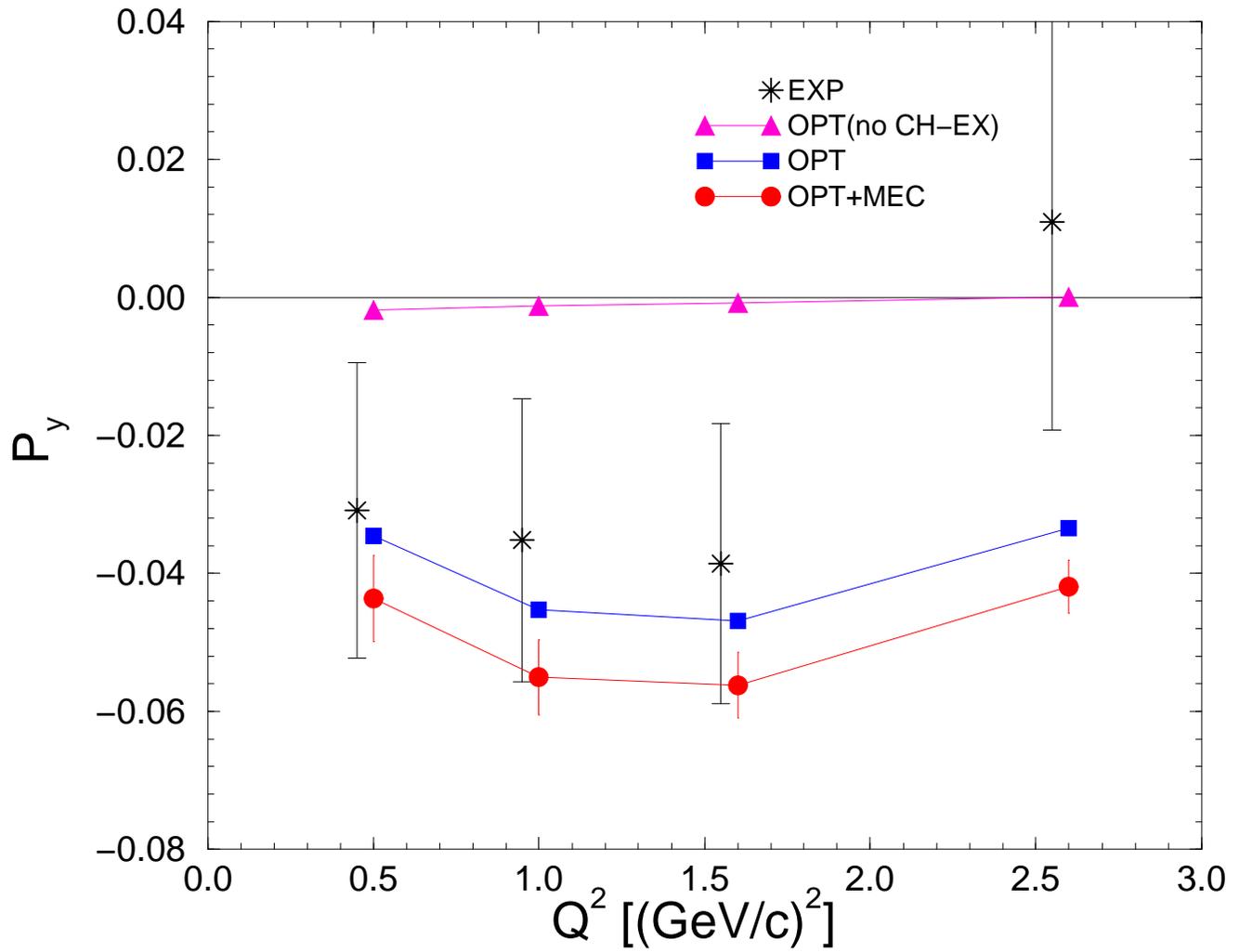}
\caption{The induced polarizations measured for $^4$He are compared to theoretical predictions,
obtained in various approximation schemes (see text for an explanation of the notation).
The solid lines are to guide the eye only.  Note that the $^4$He data have been shifted
to the left by 0.05 (GeV/c)$^2$ in order to reduce clutter.}
\label{fig:py}
\end{figure}

The matrix elements $^{(-)}\langle p+^3\!{\rm H};{\bf k}\, \sigma, \sigma_3
\mid j^\mu ({\bf q},\omega) \mid ^4\!\!{\rm He}\rangle$ are computed
with Monte Carlo (MC) techniques without making any further approximations
beyond those inherent to the treatment of FSI and nuclear electromagnetic
currents, discussed above.  The resulting theoretical predictions for the
super-ratio $R$=$(P_x^\prime/P_z^\prime)/(P_x^\prime/P_z^\prime)_{\rm PWIA}$
and for the induced polarization $P_y$ are compared with the
experimental data~\cite{Strauch03} in Figs.~\ref{fig:ratio} and~\ref{fig:py}.
The ratio of transverse to longitudinal polarizations in PWIA is proportional
to $G_{Ep}/G_{Mp}$ as obtained in the H\"ohler-Brash parameterization.
In Fig.~\ref{fig:ratio} the hydrogen data~\cite{Strauch04}
are also shown, for which, as expected, $R$ is very close to one,
except for the point at $Q^2$=1.6 (GeV/c)$^2$ (see comment above, however).

The calculated results in Figs.~\ref{fig:ratio}--\ref{fig:py}
are labeled as follows.  The curves OPT(no CH-EX) and OPT both use
one-body currents and the optical potential to describe FSI effects,
the only difference being that in the OPT(no CH-EX) calculation
the charge-exchange components $v^{c\tau}$ and $v^{b\tau}$ of
$v_T^{\rm opt}$ are ignored.  The curve labeled OPT+MEC includes
the full $v_T^{\rm opt}$ (as the curve OPT) and one- and two-body
currents.  The statistical errors associated with the MC integrations
are only shown for the OPT+MEC predictions, they are similar for the other
predictions.  Finally, the results of a calculation including one- and
two-body currents, in which the sign of the charge-exchange spin-orbit
term $v^{b\tau}$ in $v_T^{\rm opt}$ had been artificially flipped, were found
to be numerically close to those obtained in the OPT approximation, except
for $P_y$ at $Q^2$=0.5 (GeV/c)$^2$, i.e. $P_y$=$-0.0060\pm 0.0090$.  They
are not shown in Figs.~\ref{fig:ratio} and~\ref{fig:py}.

It should be stressed once more that the calculations may not describe reliably
FSI effects for the last two $Q^2$ values, since the relevant proton kinetic
energies, 0.88 GeV and 1.42 GeV, represent uncontrolled extrapolations of
the present optical model, which is fit to data up to 0.6 GeV.  For the
low $Q^2$ values, however, the $P_y$ results obtained in the OPT approximation
indicate that, while the spin-orbit terms $v^b$ and $v^{b\tau}$
may not be well constrained by the $p\, ^3$H elastic and charge exchange
differential cross sections, they seem nonetheless to be quite realistic.

The OPT+MEC calculation reproduces well the measured super-ratio $R$ at
low $Q^2$ values, and is also consistent with the measured induced polarization
$P_y$, although the experimental data for this latter quantity have rather
large systematic errors.  The charge-exchange components $v^{c\tau}$ and,
particularly, $v^{b\tau}$ in the optical potential play a crucial role---see
curves OPT(no CH-EX) and OPT in Fig.~\ref{fig:ratio}---as do two-body terms
in the electromagnetic current operator.  The inability to reproduce the
observed quenching of the super-ratio had been a persistent problem in all
earlier studies we are aware of~\cite{Udias99,Lava04,Laget04}.  Indeed,
the results of these studies are similar to those obtained here in the
OPT(no CH-EX) calculation.

In the parallel kinematics of E93-049 $P^\prime_x$ and $P^\prime_z$
are proportional, in the notation of Ref.~\cite{Picklesimer89},
to the response functions $R_{LT^\prime}^t$ and $R_{TT^\prime}^l$, respectively,
involving interference between matrix elements of charge-current and
current-current operators.  In fact, it turns out that $R_{TT^\prime}^l$=$R_T$
exactly (again, in parallel kinematics), where $R_T$ is the ordinary
transverse response.  The charge-exchange mechanism
affects both $R_{LT^\prime}^t$ and $R_{TT^\prime}^l$:
for example, at $Q^2$=0.5 (GeV/c)$^2$, $R_{LT^\prime}^t$=0.251 fm$^3$ and 0.227 fm$^3$,
and $R_{TT^\prime}^l$=0.183 fm$^3$ and 0.174 fm$^3$
in the OPT(no CH-EX) and OPT calculations, respectively.  However,
the resulting polarization transfer parameters are: $P_x^\prime$=--0.116 and --0.115,
and $P_z^\prime$=0.130 and 0.136 in the same approximations.

Two-body terms in the current operator (those in the charge operator
give tiny contributions) also affect $P_x^\prime$ and $P_z^\prime$
differently.  Both response functions $R_{LT^\prime}^t$ and $R_{TT^\prime}^l$=$R_T$
are increased by two-body current contributions, but the increase for $R_T$, about
8 \%, is twice as large as for $R_{LT^\prime}^t$ (as one would
naively expect), and therefore the ratio $P_x^\prime$/$P_z^\prime$ is
suppressed by about 4 \% with respect to that obtained with one-body
currents only.  The enhancement of $R_T$ is consistent with that calculated
for the transverse response function, measured in inclusive $^4$He$(e,e^\prime)$
scattering in quasi-elastic kinematics~\cite{Carlson00},
although it is important to emphasize that the total $p\,$$^3$H
contribution to the inclusive response involves an integral
over the missing momentum $p_m$, while here this contribution is
evaluated at a single kinematical point, namely $p_m\simeq 0$.
Lastly, among the two-body terms, the $\pi$-like and $\rho$-like
currents, derived from the isospin-dependent static part of the
AV18, and the $\Delta$-excitation current give the leading
contributions.

To conclude, the observed suppression of the super-ratio
in $^4$He is explained by FSI effects and two-body current
contributions.  In contrast to earlier suggestions made in the
literature~\cite{Strauch03}, no in-medium modification of the
proton electromagnetic form factors is needed to reproduce
the experimental data.  The present results corroborate the
conclusions derived from analyses of the Coulomb sum rule (CSR)
in few-nucleon systems~\cite{Schiavilla89}, which show
that there is no missing strength in the longitudinal response of
these nuclei when the free-proton electric form factor is
used.  The CSR situation for medium-weight nuclei
remains controversial to this day~\cite{Jourdan96,Morgenstern01},
although there are rather strong indications that even there
no quenching of longitudinal strength is observed~\cite{Jourdan96,Sick04}.
Therefore, the quark-meson coupling model of nucleon and nuclear
structure~\cite{Lu99}, which leads to the notion of
medium-modified nucleon form factors, seems to be at variance with
a number of experimental observations.  It is interesting to note that
this notion is not an inevitable consequence of the quark substructure
of the nucleon.  For example, a recent study~\cite{Paris00} of the two-nucleon 
problem in a flux-tube model of six quarks interacting via single
gluon and pion exchanges suggests that the nucleons retain
their individual identities down to very short separations, with
little distortion of their substructures.

We thank S.\ Strauch and P.\ Ulmer for correspondence in reference to various
aspects of the experiment, and V.R.\ Pandharipande and I.\ Sick for a
critical reading of the manuscript.
The work of R.S.\ was supported by DOE contract DE-AC05-84ER40150
under which the Southeastern Universities Research Association (SURA)
operates the Thomas Jefferson National Accelerator Facility.
All the calculations were made possible by grants
of computing time from the National Energy Research Supercomputer
Center.
%
%
%

%
%
%
%
\newpage
\newpage
\end{document}